\documentclass[aps,prl,showpacs,twocolumn,amsmath,amssymb]{revtex4}

\usepackage[english]{babel}
\usepackage{latexsym}
\usepackage{graphics,psfrag}
\usepackage{epsfig}
\usepackage{color}

\def\be{\begin{equation}}
\def\ee{\end{equation}}
\def\bea{\begin{eqnarray}}
\def\eea{\end{eqnarray}}
\def\bi{\begin{itemize}}
\def\ei{\end{itemize}}
\def\bin{\begin{enumerate}}
\def\ein{\end{enumerate}}

\newcommand{\vect}[1]{\mathbf{#1}}

\begin{document}
\title{Matter-wave analog of an optical random laser}

%%%%%%%%%%%%%%%%%%%%%%%%%%%%%%%%%%%%%%%%%%%%%%%%%%%%%%%%%%%%%%%%%%%%%%%%%%%%%%%
\author{Marcin P\l{}odzie\'n}
\affiliation{
Instytut Fizyki imienia Mariana Smoluchowskiego and
Mark Kac Complex Systems Research Center, 
Uniwersytet Jagiello\'nski, ulica Reymonta 4, PL-30-059 Krak\'ow, Poland}

\author{Krzysztof Sacha}
\affiliation{
Instytut Fizyki imienia Mariana Smoluchowskiego and
Mark Kac Complex Systems Research Center, 
Uniwersytet Jagiello\'nski, ulica Reymonta 4, PL-30-059 Krak\'ow, Poland}

\date{\today}

\begin{abstract}
The accumulation of atoms in the lowest energy level of a trap and
the subsequent out-coupling of these atoms 
is a realization of a matter-wave analog of a conventional optical laser.
Optical random lasers require materials that provide optical gain but, 
contrary to conventional lasers, the modes are determined by multiple 
scattering and not a cavity. We show that a Bose-Einstein condensate can be
loaded in a spatially correlated disorder potential prepared in such a way 
that the Anderson localization phenomenon operates as a band-pass filter. 
A multiple scattering process selects atoms with certain momenta and determines laser 
modes which represents a matter-wave analog of an optical random laser.
\end{abstract}


\maketitle
%%%%%%%%%%%%%%%%%%%%%%%%%%%%%%%%%%%%%%%%%%%%%%%%%%%%%%%%%%%%%%%%%%%%%%%%%%%%%%%

Conventional optical lasers require two ingredients: material that 
provides optical gain and an optical cavity responsible for coherent
feedback and selection of resonant laser modes. However, it is
also possible to achieve laser action without the optical cavity
provided the gain material is an active medium with disorder \cite{wiersma08}. 
Forty years ago
Letokhov analyzed a light diffusion process with amplification and predicted that 
gain could overcome loss if the volume of a system exceeded a critical 
value \cite{letokhov67}. 
Random lasing (i.e. light amplification in disordered gain media), 
achieved in a laboratory in the 1990's, 
attracts much experimental attention and 
offers possibilities for interesting applications 
\cite{lawandy94,wiersma96,cao98,frolov99,wiersma00,wiersma08}. 
Theoretical understanding of this phenomenon is still imperfect. 
Although the Letokhov model of diffusion with gain is useful in predicting 
certain properties of random lasers, it neglects coherent phenomena. There are
various theoretical models of random lasing but 
it is widely accepted that 
%certainly 
interference 
in a multiple scattering process determines the spatial and spectral mode 
structure of a random laser \cite{wiersma08}.

Bose-Einstein condensation (BEC) of dilute atomic gases is a macroscopic 
accumulation of atoms in the lowest energy level of a trap when the temperature of
the gas decreases \cite{anderson95}. This tendency of occupying a single state  through the mechanism of stimulated
scattering of bosons is an analog of mode selection in optical lasers due 
to the stimulated emission of photons. 
Gradual release of atoms from a trapped BEC allows for 
the realization of a matter-wave analog of a conventional optical laser
\cite{mewes97,hagley99,bloch99,cennini03,guerin06,tomek08}. 
The atom trap is an analog of the optical cavity. The lowest mode of the trap
is a counterpart of an optical resonant mode. In conventional 
optical lasers the output coupler is usually a partially transmitting 
mirror. In atom lasers it involves, for example, 
a change of the internal state of the atom by means of a radio-frequency transition. 
%Quasi-continues and pulsed atom lasers have been realized in laboratory
%and properties of the atom lasers, analogous to the properties of 
%the optical counterparts, have been demonstrated experimentally
%\cite{mewes97,hagley99,bloch99,cennini03,guerin06,tomek08,
%andrews97,hagley99a,kozuma99,inouye99,deng99}. 

In the present letter we propose the realization of a matter-wave analog of an
optical random laser. Suppose
the BEC of a dilute atomic gas has been achieved in 
a trapping potential. That is, we begin with 
the accumulation of atoms in a single mode of the resonator 
(i.e. the lowest eigenstate of the trap). Then, let us turn off 
the trap and turn on a weak disorder potential. 
Starting with a BEC we have a guarantee that the 
disorder medium is {\it pumped} with coherent matter-waves. 
We would like to raise the question of whether it is possible to prepare spatially
correlated disorder potential in such a way that narrow peaks can be observed 
in the spectrum of atoms that are able to 
leave the area of the disorder potential? In other words:
if the multiple scattering of atoms in a disorder medium can lead to a selective
spectral emission of matter-waves from the medium? 

In cold atom physics a disorder potential can be realized by means of an optical
speckle potential \cite{schulte05,clement06}. 
Transmission of coherent light through a diffusing plate
leads to a random intensity pattern in the far field. 
% from the plate. 
Atoms experience the presence of the radiation as an external potential 
$V(\vect r)\propto \chi|E(\vect r)|^2$ proportional to the intensity of the
light field $E(\vect r)$ and atomic polarizabitlity $\chi$ whose sign 
depends on the detuning of the light frequency from the atomic resonance.
Diffraction from the diffusive plate onto the location of atoms
determines correlation functions of the speckle potential. We assume
that the origin of the energy is shifted so that $\overline{V(\vect r)}=0$ where
the overbar denotes an ensemble average over disorder realizations. Standard
deviation of the speckle potential $V_0$ measures the strength of the disorder.

Let us begin with a one-dimensional (1D) problem. In a weak disorder potential 
atoms with $k$-momentum undergo multiple scattering, 
diffusive motion and finally localize with an exponentially decaying
density profile due to the Anderson localization process, provided that the 
system size exceeds the localization length \cite{anderson58,lee85,tiggelen99}. 
Taking the Born approximation to the second order in the potential strength, 
the inverse of the localization length is \cite{lifshits88}
%$l_{loc}^{-1}=\sqrt{2\pi}(mV_0/\hbar^2k)^2{\cal P}(2k)$,
$l_{loc}^{-1}=(mV_0/\hbar^2k)^2{\cal P}(2k)$,
where the Fourier transform 
of the pair correlation function of the speckle potential is
\be
{\cal P}(k)=\int \frac{dq}{2\pi}\gamma(q)\gamma(k-q),
\label{pk}
\ee
and $\gamma(k)=\int dz \tilde\gamma(z)e^{-ikz}$ 
is the Fourier transform of the complex degree of coherence  
%$\gamma(z)=\overline{E^*(z+z')E(z')}/\overline{|E(z)|^2}$.
$\tilde\gamma(z)=\overline{E^*(z+z')E(z')}/\overline{|E(z)|^2}=\int dy {\cal A}(y)e^{izy/\alpha}/\int dy {\cal A}(y)$. 
${\cal A}(y)$ describes the aperture of the optics and 
$\alpha$ is a constant dependant on the wavelength of the laser radiation and
the distance of the diffusive plate from the atomic trap.
In Ref.~\cite{billy08}, where the experimental realization of 
the Anderson localization of matter-waves is
reported (see also \cite{roati08}), 
a simple Heaviside step function ${\cal A}(z)=\Theta(R-|z|)$ describes 
the aperture. The corresponding
$\gamma(k)=\pi\sigma_R\Theta(1-|k\sigma_R|)$, where
$\sigma_R=\alpha/R$ is the correlation length of the speckle potential. 
Consequently 
the power spectrum (\ref{pk}) decreases linearly 
and becomes zero for $|k|\ge 2/\sigma_R$. Thus, the Born approximation predicts 
an effective mobility edge at $|k|=1/\sigma_R$, i.e. atoms with larger 
momenta do not localize \cite{sanchez07,billy08} 
(actually higher order calculations \cite{lugan09} show  
they do localize but with very large localization lengths, much larger than the
system size in the experiment). 
Hence, neglecting atom interactions, if the width of the initial 
atom momentum distribution exceeds the mobility edge
particles at the tail of the distribution avoid the Anderson localization 
and may leave the disorder area.

%%%%%%%%%%%%%%%%%%%%%
\begin{figure}%[h]
\centering
\includegraphics*[width=0.95\linewidth]{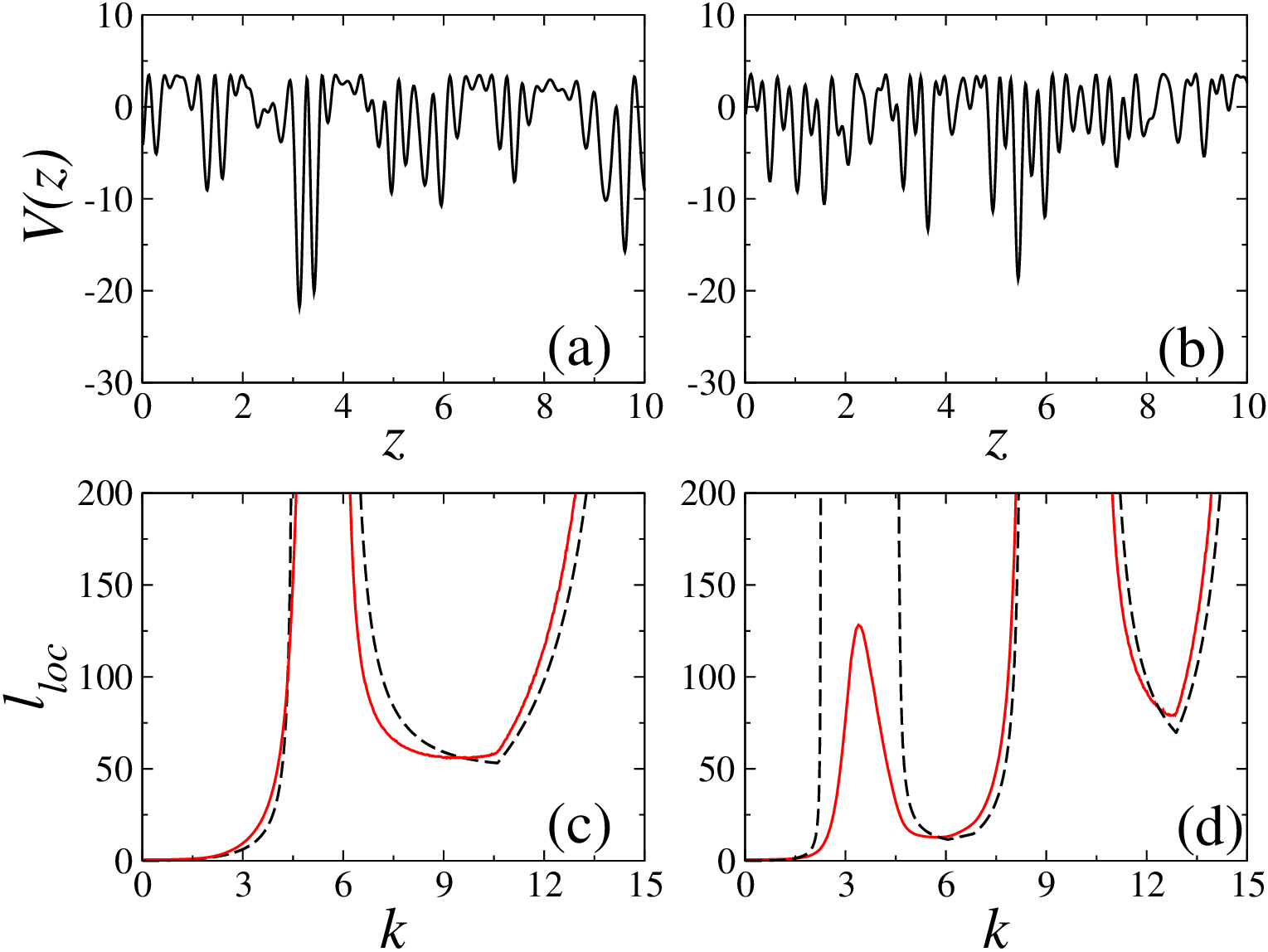}
\caption{(Color online) 
Examples of the speckle potential (top panels) and the corresponding
localization length (bottom panels) obtained within the Born approximation
(dashed black curves) and numerically in the transfer-matrix calculations (solid
red curves).
Panels (a) and (c) show the results for the single obstacle in the diffusive plate 
where $\sigma_R=0.066$ (0.31~$\mu$m), $\sigma_R/\sigma_\rho=0.4$ and $V_0=3.5$. Panels (b) and (d)
correspond to the case of two obstacles with the same $\sigma_R$ and $V_0$ 
but $\sigma_R/\sigma_\rho=0.7$ and $\sigma_R/\sigma_\zeta=0.1$. 
The results are shown for
rubidium-87 atoms. Red detuning of the laser radiation from the atomic 
resonance is assumed. All values are presented in the harmonic oscillator 
units, i.e. energy $E_0=\hbar\omega$ and length
$l_0=\sqrt{\hbar/m\omega}$ where $\omega/2\pi=5.4$~Hz.
%Note that while the potentials in (a) and (b) do not behave very differently
%the behaviour of the corresponding $l_{loc}$'s is very different.
}
\label{one}
\end{figure}
%%%%%%%%%%%%%%%%%%%%%

Let us modify the experiment reported in Ref.~\cite{billy08} by
introducing an obstacle at the center of the diffusive plate so that 
the aperture is now described by 
${\cal A}(z)=\Theta(R-|z|)-\Theta(\rho-|z|)$ where 
$\rho<R$. It implies that
\be
\gamma(k)=\pi\left(\frac{1}{\sigma_R}-\frac{1}{\sigma_\rho}\right)^{-1}[
\Theta(1-|k\sigma_R|)-\Theta(1-|k\sigma_\rho|)],
\ee
where $\sigma_\rho=\alpha/\rho$. If the size of
the obstacle $\rho>R/3$, interference of light passing through such a 
{\it double-slit } diffusive
plate creates a peculiar speckle potential. That is, the power spectrum 
(\ref{pk}) disappears for $|k|\ge 2/\sigma_R$ as previously but it is
also zero for
$\frac{1}{\sigma_R}-\frac{1}{\sigma_\rho}<|k|<\frac{2}{\sigma_\rho}$.
Thus, according to the Born approximation there is a momentum interval where 
the localization length diverges. It implies that the
Anderson localization process is able to operate as a band-pass filter 
letting particles with specific momenta leave the region of the disorder. 
Detection of escaping atoms should reveal a peak in the momentum 
spectrum corresponding to the interval where the localization length diverges. 
%In Fig.~\ref{one} we present an example of 
%the speckle potential and the localization length obtained within 
%the Born approximation and numerical values from a transfer-matrix calculation.

%%%%%%%%%%%%%%%%%%%%%
\begin{figure}%[h]
\centering
\includegraphics*[width=0.9\linewidth]{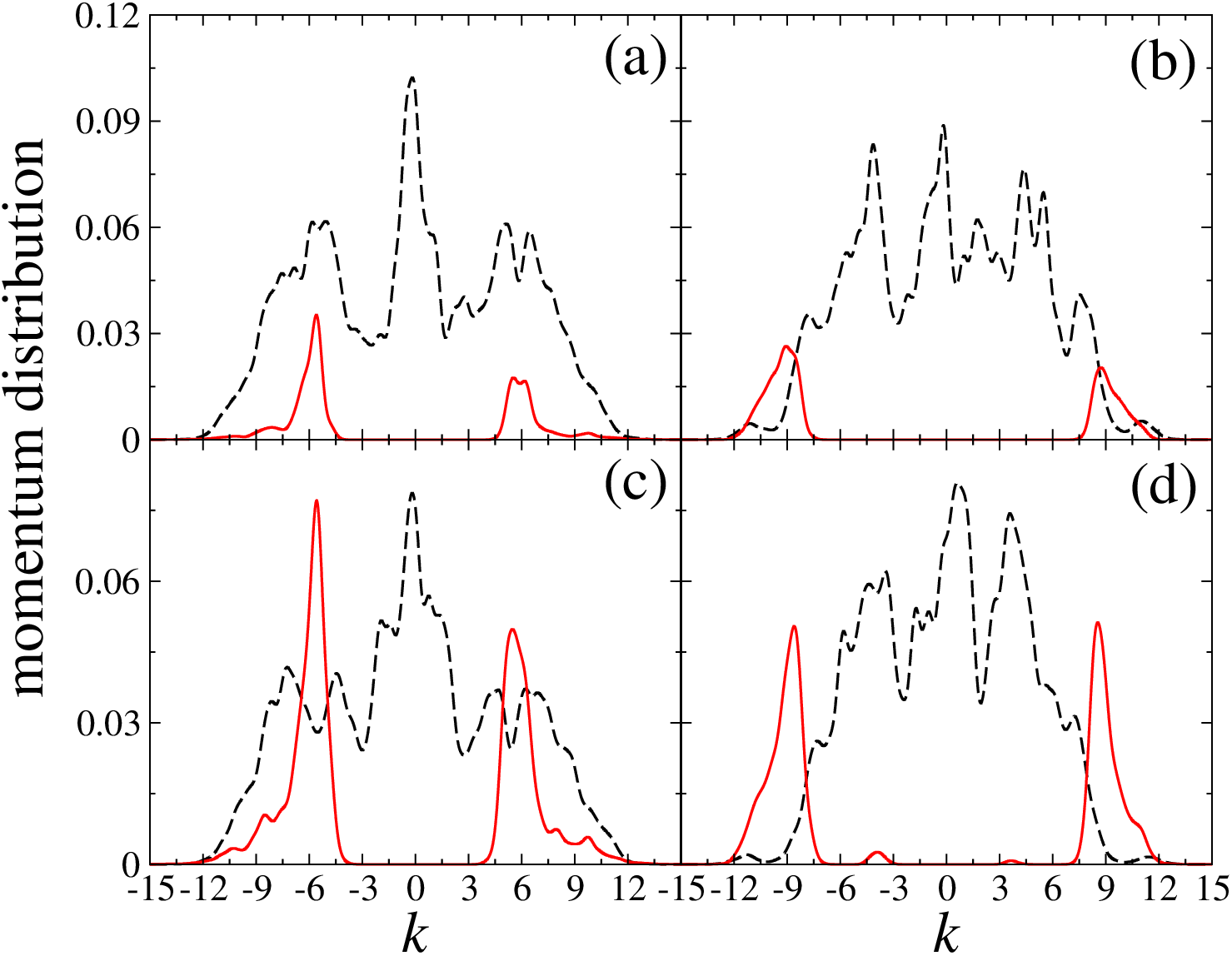}
\caption{(Color online) 
Momentum distributions of the atoms localized in the disorder 
potential (dashed black lines) and the atoms which escaped from the disorder area 
(solid red lines). Panels (a) and (c) show the results for a single experimental 
realization of the disorder with parameters as in Fig.~\ref{one}a,c 
while panels (b) and (d) are related to the parameters as in Fig.~\ref{one}b,d.
Panel (a) corresponds to the evolution time $t=100$ (2.9~s), panel (b) to 
$t=70$ (2~s), panels (c) and (d) to $t=200$ (5.7~s). Fraction of atoms that 
escaped the disorder region is about 9\% in (a) and (b), and 20\% in (c) and (d).
In order to take into account experimental resolution all data have been 
convoluted with Gaussian of $\Delta k=0.3$ width. 
All values are presented in the harmonic oscillator 
units, i.e. energy $E_0=\hbar\omega$, length
$l_0=\sqrt{\hbar/m\omega}$ and time $t_0=1/\omega$ where $\omega/2\pi=5.4$~Hz.}
\label{two} 
\end{figure}
%%%%%%%%%%%%%%%%%%%%%

Introducing two (or more) obstacles in the diffusive plate we can increase 
the number of momentum intervals with diverging $l_{loc}$. 
In Fig.~\ref{one} we present examples of the speckle potentials in 
the single obstacle case and the case of 
two obstacles located symmetrically around 
the plate center. In the latter case the aperture is described by 
${\cal A}(z)=\Theta(R-|z|)-\Theta(\rho-|z|)+\Theta(\zeta-|z|)$ where
$\zeta<\rho$ and 
\bea
\gamma(k)&=&
\pi\left(\frac{1}{\sigma_R}-\frac{1}{\sigma_\rho}+\frac{1}{\sigma_\zeta}\right)^{-1}
[\Theta(1-|k\sigma_R|) \cr &&-\Theta(1-|k\sigma_\rho|)
+\Theta(1-|k\sigma_\zeta|)],
\eea
with $\sigma_\zeta=\alpha/\zeta$.
The figure also presents localization lengths obtained numerically in 
the transfer-matrix calculation \cite{lugan09} that confirms the Born predictions. 

To simulate an experiment, we follow the parameters used 
in Ref.~\cite{billy08} where Anderson localization of matter-waves has been
observed. We assume that a BEC of $N=1.7\cdot 10^4$ 
rubidium-87 atoms is initially prepared in a quasi-1D harmonic trap  
with longitudal and transverse frequencies $\omega/2\pi=5.4$~Hz and 
$\omega_\perp/2\pi=70$~Hz, respectively. 
In the following we adopt the harmonic oscillator units:
$E_0=\hbar\omega$, $l_0=\sqrt{\hbar/m\omega}$ and $t_0=1/\omega$ for energy,
length and time, respectively. 
When the trapping potential is turned off and the speckle potential
is turned on the expansion of the atomic cloud is initially dominated by the
particle interactions until the density drops significantly and the atoms start
feeling only the disorder potential. This initial stage of the
gas expansion sets the momentum distribution of the atoms which may be
approximated by an inverted parabola with an upper cut-off 
$k_{max}=2\sqrt{\mu}=12.7$ where $\mu$ is the initial chemical potential of 
the system \cite{sanchez07,miniatura09}. 

The disorder potentials we choose are attainable in the
experiment reported in Ref.~\cite{billy08}, i.e. 
they extend 862 units (4~mm) along the $z$ direction with
the correlation length $\sigma_R=0.066$ (0.31~$\mu$m). 
We consider the potentials obtained by introducing one or two 
obstacles in the diffusive plate that are presented in Fig.~\ref{one}. 
%obstacles introduced in the diffusive plate leads to 
%$\sigma_R/\sigma_\rho=0.7$ and $\sigma_R/\sigma_\zeta=0.1$, 
%the strength of the disorder $V_0=3.5$ (i.e. $V_0/\mu=0.086$). An
%example of the potential is presented in Fig.~\ref{one}b and the corresponding
%localization length is plotted in Fig.~\ref{one}d. 
If the atomic cloud starts at the center of the disorder potentials
%that extends 862 along longitudal direction 
and if the cut-off of 
the momentum distribution is $k_{max}\lesssim 13$ then we may expect that 
time evolution leads to emission of atoms with $|k|\approx 5.5$ in the single
obstacle case (cf. Fig.~\ref{one}a,c) and with $|k|\approx 9$ in the case of two
obstacles (cf. Fig.~\ref{one}b,d).
In the latter case we may expect also a small leakage of atoms with 
$|k|\approx 3.5$. For $|k|\approx 3.5$ the localization length shown in 
Fig~\ref{one}d reaches locally 
a maximum value of $l_{loc}\approx 120$.
%This is because 
Before the particle interactions become negligible the gas spreads over 
a significant range of the disorder region. Therefore 
the Anderson localization is not 
able to diminish completely the leakage of atoms with $|k|\approx 3.5$.
%The localization length corresponding to local maximum at $k\approx 4$ is comparable
%to the distance from the center to the boundary of the
%disorder potential. Thus, a small leakage of atoms with $k\approx 4$ is
%also expected. 

Starting with the ground
state of the stationary Gross-Pitaevskii equation
\cite{mewes97,hagley99,bloch99,cennini03,guerin06,tomek08,schulte05} 
in the presence of 
the harmonic trap, we integrate the time-dependent Gross-Pitaevskii
equation when the trap is turned off and a disorder is turned on.
Figure~\ref{two} shows momentum distributions of atoms that
escaped from the disorder region and those that are localized 
for the disorder potentials corresponding to Fig.~\ref{one}
at different moments in time. The expected selective spectral 
emissions of atoms are apparent in the figure. Interestingly 
in Fig.~\ref{two}d, i.e. for longer evolution time,
small peaks around $|k|\approx 3.5$ become visible. 
%If we simulate evolution of a BEC for the parameters of the potential presented
%in Fig.~\ref{one}a and \ref{one}c

%To detect the signatures of atom random lasing in experiments one can 
%either measure momenta of escaping atoms or momentum distribution of the 
%atoms that
%localize in the disorder potential. The latter option seems to be quite simple. 
%Indeed, turning off the disorder the initially localized atoms 
%expand and their density reflects the initial momentum distribution after 
%a short time.

%%%%%%%%%%%%%%%%%%%%%
\begin{figure}%[h]
\centering
\includegraphics*[width=0.9\linewidth]{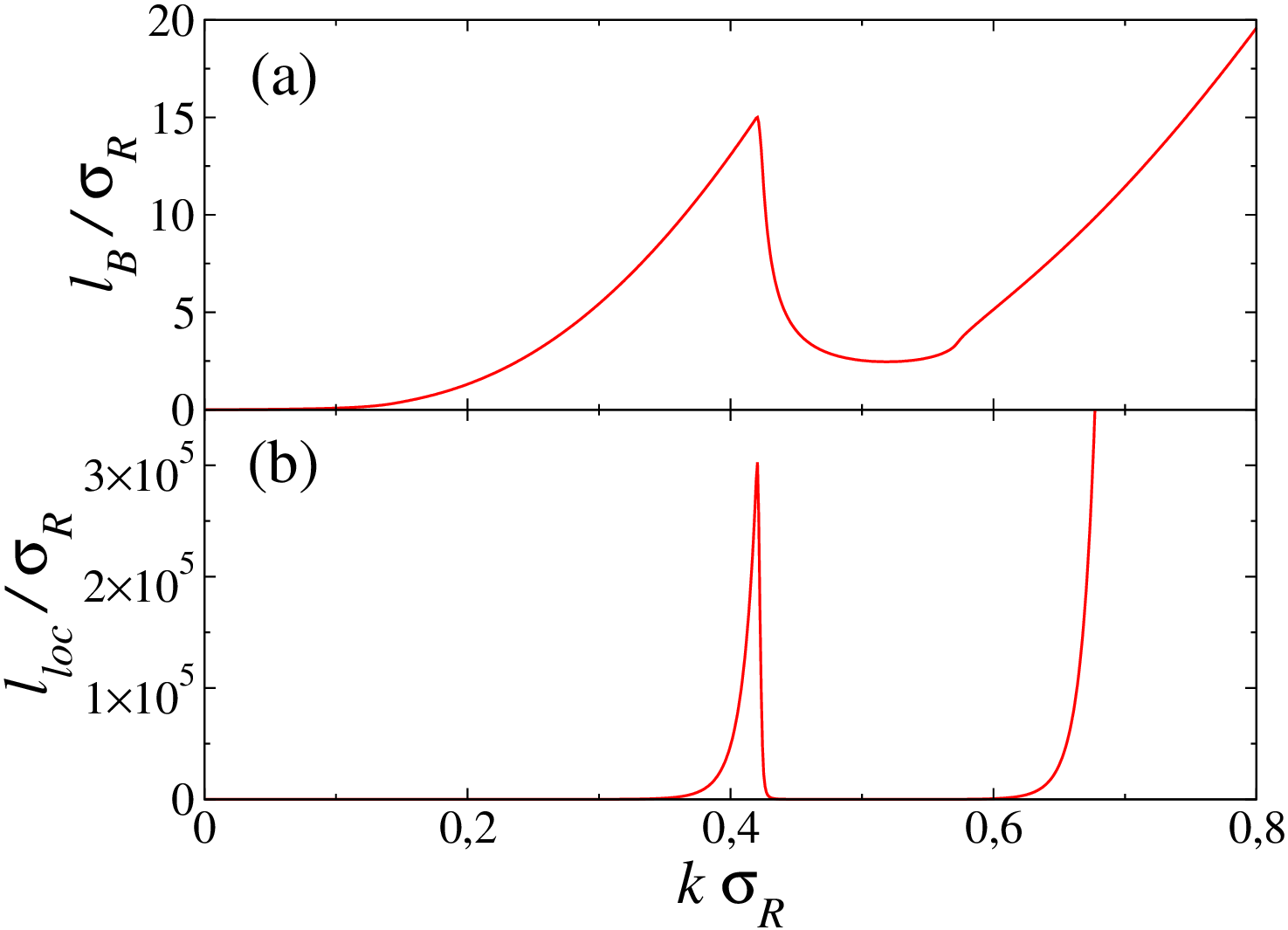}
\caption{(Color online) 
The Boltzmann transport mean-free path (a) and the localization length (b) 
for atoms in the 2D speckle potential created by transmission of a laser
beam through the circularly shaped diffusive plate with the obstacle in the 
form of a ring, i.e. the aperture of the optics is described by 
$\Theta(R-|\vect r|)-\Theta(\rho-|\vect r|)+\Theta(\zeta-|\vect r|)$
with $\sigma_R/\sigma_\rho=0.99$ and $\sigma_R/\sigma_\zeta=0.15$.
%$\rho=0.99R$ and $\zeta=0.15R$. 
The potential strength corresponds to $\eta=0.15$.
The quantities presented in the figure are dimensionless.
% [see (\ref{2dlb})].
}
\label{three}
\end{figure}
%%%%%%%%%%%%%%%%%%%%%

Finally let us consider a possibility of the realization of an atom analog of
an optical random 
laser in 2D. The Boltzmann transport mean-free path $l_B$ is the characteristic
spatial scale beyond which memory of the initial direction of the 
particle momentum is lost. In 2D $l_{loc}= l_Be^{\pi kl_B/2}$ and thus
the localization length is much larger than 
$l_B$ which is in contrast to 
the 1D case where these two quantities are nearly
identical $l_{loc}=2l_B$ \cite{tiggelen99}. 
For the circularly shaped diffusing plate 
with the radius $R$ the classical transport 
mean-free path \cite{kuhn05,miniatura09}, 
to the second order in the potential strength, reads
\be
\frac{1}{l_B}=\frac{\eta^2}{k\sigma_R^2}
%\frac{V_0^2m^2\sigma_R^2}{k\hbar^4}
\int \frac{d\phi}{2\pi}(1-\cos\phi){\cal
P}\left(2k\sigma_R\sin\frac{\phi}{2}\right),
\label{2dlb}
\ee 
where $\eta=V_0/E_\sigma$ is 
the ratio of the potential strength and correlation energy
$E_\sigma=\hbar^2/(m\sigma_R^2)$ with $\sigma_R=\alpha/R$.
The power spectrum ${\cal P}(k)$ of the optical speckle potential 
disappears for $k\ge 2/\sigma_R$. Nevertheless,
the $l_B$ (and consequently also $l_{loc}$) is always finite.
% for all momenta because even fast atoms can be slightly deflected by 
%a smooth speckle potential. 
In the bulk 2D system, an initially prepared atomic wave-packet 
follows a diffusive motion at short time but eventually the dynamics slow 
down and freeze due to the Anderson localization process 
\cite{kuhn05,miniatura09,vincent10}. 

By introducing obstacles in the diffusive plate we are able to shape the
power spectrum of the speckle potential. On one hand the fact that 
${\cal P}(k)$ may disappear at certain momentum intervals does not mean
divergence of the corresponding transport mean-free path (\ref{2dlb}). 
On the other hand any non-monotonic 
behaviour of $l_B(k)$ is dramatically amplified in the behaviour of 
$l_{loc}(k)$ because the localization length is an exponential function of
$l_B$. In Fig.~\ref{three} we present an example related to the obstacle
in the form of a ring, i.e. the aperture of the optics is described by
${\cal A}(\vect r)=\Theta(R-|\vect r|)-\Theta(\rho-|\vect r|)+\Theta(\zeta-|\vect r|)$.
% with $\rho=0.99R$ and $\zeta=0.15R$ and the potential strength 
% corresponding to $\eta=0.2$. 
At $k\sigma_R\approx 0.4$ both $l_B$ 
and $l_{loc}$ shows a maximum. However, while the transport mean-free path 
changes by only a few in the neighboring region 
the localization length changes by four orders of magnitude. 
If the width of the momentum distribution of a BEC loaded
in such a disorder potential
is smaller than $0.6/\sigma_R$ and the radius of the disorder medium is greater
than $10^3\sigma_R$ but
less than $10^5\sigma_R$ the multiple scattering 
process leads to an isotropic emission
of atoms with $k\approx 0.4/\sigma_R$.

We have outlined a proposal for the realization of a matter-wave analog of an 
optical random laser. Spatially correlated disorder potential for atoms 
with a peculiar pair correlation function can be created by transmitting 
a laser beam through a diffusive plate with obstacles. The resulting Anderson
localization length reveals non-monotonic behaviour as a function of particle
momentum. It allows for filtering momenta of particles that leave the area of
the disorder, if the size of the disorder medium is suitably chosen. The 
disorder medium is assumed to be initially loaded with a BEC 
which guarantees that the matter-waves emitted from the medium are
coherent. We have restricted ourselves to the 1D and 2D cases but the atom
analog of an optical
random laser can be also anticipated in 3D. 
In 3D the Ioffe-Regel criterion discriminates between waves that are Anderson localized
($kl_B\lesssim 1$) or not \cite{tiggelen99,kuhn07}. Thus, a spatially correlated 
disorder potential for which the Ioffe-Regel
criterion is not fulfilled for specific momenta should allow for selective 
emission of matter-waves 
%and the realization of an atom random laser 
in 3D.

Our proposal is directly applicable to atomic matter-wave experiments. 
From the point of view of the optical random lasers our analysis is not 
complete because it is restricted to passive random materials without gain.
There is an interesting question whether a disorder with properties similar 
to those analyzed here play a role in optical random lasers and which modes 
are important when the gain is included in a system. 

The non-monotonic behaviour of the localization length 
results in the appearance of a multiple effective mobility edge if a disorder
system is finite. Wave transport is then unusual and interesting on its own. 
A shallow non-monotonical bahaviour of the Anderson
localization length versus energy has been 
observed also in a classical wave system,
see Ref.~\cite{chorwat}.

%%%%%%%%%%%%%%%%%%%%%%%%%%%%%%%%%%%%%%%%%%%%%%%%%%%%%%%%%%%%%%%%%%%%%%%%
We are grateful to D. Delande for encouraging discussion.
 and to R. Marcinek and
 J. Zakrzewski for critical reading of the manuscript.
This work is supported by the Polish Government within research projects
 2009-2012 (MP) and 2008-2011 (KS).

{\it Note added:} After submission of this article, we became aware of a related
theoretical study \cite{laurent11}.

%%%%%%%%%%%%%%%%%%%%%%%%%%%%%%%%%%%%%%%%%%%%%%%%%%

\end{document}